\def\ie{{\it i.e.}}

\def\inv{{\cal I}}
\def\sym{{\cal S}}
\def\gauge{{\cal G}}
\def\diff{{\cal D}}
\def\conn{{\cal C}}
\def\met{{\cal M}}

\def\lie{{\cal L}}

\null

\centerline{\bf Mixing internal and spacetime transformations: }
\centerline{\bf some examples and counterexamples}
\smallskip
\centerline{Roberto Percacci}
\centerline{SISSA, via Beirut 4, I-34014 Trieste, Italy}
\centerline{INFN, Sezione di Trieste, Italy}

\midinsert
\narrower
Abstract: this note addresses the question whether in a gauge theory 
coupled to gravity internal and spacetime transformation can be mixed.
It is shown that if the VEV of the gauge field is flat, the
symmetry group is always a product of internal and spacetime symmetries.
On the other hand, if the VEV of the gauge field is not flat
it is impossible to properly define the notion of a ``spacetime'' transformation;
as a consequence, if the symmetry group is nontrivial, mixing generically occurs.
\endinsert

\bigskip


The celebrated Coleman-Mandula (CM) theorem [1]
asserts that, under rather general conditions,
the symmetries of the S-matrix must be the product of
the Poincar\'e group and some internal symmetry group.
It forbids mixing spacetime and internal {\it symmetries}.
This theorem is sometimes misunderstood as forbidding
any mixing between internal and spacetime {\it invariances}.
The mistake here lies in applying it to transformations
that are not symmetries of the $S$ matrix.
For example, a Yang-Mills theory in flat 
space is invariant under Poincar\'e transformations
and under local {\it gauge} transformations, and these two invariance groups do not commute.
Such mixing between spacetime 
and internal transformations is not forbidden by the CM theorem.

On the other hand, since the existence of a Poincar\'e subgroup is one
of the hypotheses of the theorem, it is sometimes said that a quantum field theory
in a curved background metric will not be subject to the same restrictions.
While the CM theorem itself does not apply directly to these cases, 
some generalization thereof could still lead to similar conclusions.
In fact I will show below that in a large class of examples
where the CM theorem cannot be applied,
spacetime and internal symmetries still do not mix.

Spacetime transformations feature prominently in the
theory of gravitation, so a generalization of the CM theorem
to that context would be very useful, especially if it could
put restrictions on possible ways of unifying 
gravity with other interactions.
However, the proof of the theorem relies heavily on 
the use of mathematical tools that are peculiar to flat space.
It is not immediately clear how to generalize it to other situations.
For this reason, in this note we will discuss the possible mixing between internal
and spacetime transformations starting from an entirely different set of hypotheses.
We will restrict ourselves to the case of gauge theories,
meaning theories whose dynamical variables are a gauge field $A_\mu$
with values in the Lie algebra of a group $G$,
matter fields $\psi$ carrying linear representations $\rho$ of $G$,
and a metric on spacetime, $g_{\mu\nu}$.
Insofar as our present understanding of fundamental interactions
is based on gauge theories, this is not much of a restriction.
Furthermore, non-gauge theories can be seen as special cases
where $A_\mu$ is non-dynamical and flat 
and similarly one can switch off gravity by declaring $g_{\mu\nu}$
non-dynamical and flat.

For the subsequent discussion it is important to define precisely
what is meant by {\it symmetry group} of a theory.
In general, the symmetries are not entirely determined by the Lagrangian:
they depend also on the properties of the vacuum state,
more precisely on the Vacuum Expectation Value (VEV) of the fields.
For example a scalar theory can exist in different phases
characterized by different VEVs, and each phase has a different symmetry group.
In a gauge theory the situation is more complicated.
To see why, let us introduce some terminology.
We will call $\inv$ the ``invariance group of the theory'',
namely the transformations of the fields that leave the action invariant.
It consists of ``Yang-Mills'' transformations parametrized by some function
$g(x)$ with values in $G$ (possibly a constant) and diffeomorphisms 
$x'=f^{-1}(x)$
\footnote{($^1$)}{Throughout this note we will use the active point of view;
all transformations change the fields leaving the coordinates fixed.}
.
In infinitesimal form, if $g=1+\epsilon$,
$$
\delta_\epsilon A_\mu=D_\mu\epsilon\ ;\ \ \ \delta_\epsilon\psi=-\rho(\epsilon)\psi\ .\eqno(1)
$$
Infinitesimal diffeomorphisms $x^{\prime\mu}(x)=x^\mu-\xi^\mu(x)$
are parametrized by vectorfields
$\xi^\mu$ and the variation of any field $\phi$ (here including also
the gauge fields $A_\mu$ and the metric $g_{\mu\nu}$)
under such transformation is given by the Lie derivative
$$
\delta_\xi\phi=\lie_\xi\phi\ .\eqno(2)
$$

From a physical standpoint,
we will then distinguish two classes of invariances.
If a transformation: (g1) leaves the action invariant and
(g2) is such that the transformed fields
are {\it physically indistinguishable} from the original ones, 
then we will call it a ``gauge invariance''.
We denote $\gauge\subset\inv$ the group of gauge invariances.
We will see that it is a normal subgroup of $\inv$. 
The presence of a gauge invariance means that the kinematical description of the theory
is redundant, because it contains unphysical degrees of freedom.
On the other hand if a transformation: (s1) leaves the action invariant,	
(s2) is such that the transformed fields
{\it can be physically distinguished} from the original ones,
and (s3) it leaves the vacuum invariant, we will call it a ``symmetry''.
Unlike gauge transformations, symmetries by definition depend 
on the VEVs - a set of classical fields.

A convenient way of studying gauge theories is the background field method.
One splits the metric and the gauge field into classical
backgrounds $g^{(0)}_{\mu\nu}$ and $A^{(0)}_\mu$
and quantum fluctuations $h_{\mu\nu}$ and $a_\mu$.
The backgrounds are identified a posteriori with the VEVs of the respective fields,
so that $\langle h_{\mu\nu}\rangle=0$ and $\langle a_\mu\rangle=0$.
Strictly speaking this procedure cannot be correct because
the dynamics of the theory is gauge invariant and it cannot select
a particular representative in the gauge equivalence class
\footnote{($^2$)}{I wish to thank Abhay Ashtekar for a discussion
of this point.}.
For example, in the case of Yang--Mills theories this description contradicts 
the Elitzur theorem [2], which says that only gauge invariant operators
can have nonzero VEV. See also [3] for a related discussion.
In spite of this, in the first part of this paper we will stick to the popular 
background field terminology and identify ``the vacuum'' with a particular choice of
classical fields $(A^{(0)}_\mu,g^{(0)}_{\mu\nu})$.
We call $\sym^{(0)}\subset\inv$ the group of symmetries corresponding to these classical fields.
We will discuss the groups $\inv$, $\gauge$ and $\sym^{(0)}$ in a number of cases,
finding both examples and counterexamples to the
possibility of mixing internal and spacetime transformations.
It will be shown that the structure of the symmetry group does not depend on the particular choice
of representative in a gauge equivalence class.
Since there may still be doubts about the correctness of this procedure,
we will then briefly discuss an alternative formulation of the problem in which 
gauge invariance is not broken.
In this formulation the vacuum will be described by a gauge equivalence class
of classical fields, and it will be shown that the same results are obtained.

As a warmup, let us begin by discussing the case when the VEVs are
$$
A^{(0)}_\mu=0\ ;\ \ \ g^{(0)}_{\mu\nu}=\eta_{\mu\nu}\ .\eqno(3)
$$ 
The theory describes gauge bosons, matter fields and gravitons
propagating in flat space.
This is the domain in which one expects the CM theorem to apply.

As is well-known, Yang-Mills fields can be interpreted geometrically
as connections in a principal bundle $P$, while matter fields are
sections of associated bundles.
For our purposes this has the advantage that instead of talking of the
action of $\inv$ on the fields, given by (1,2), we can talk about
its action on $P$, which is easier to visualize.
In the case at hand $P$ is trivializable, meaning that it is
diffeomorphic to a product $M\times G$,
($M$ is spacetime), so we can coordinatize $P$ by
pairs $(x,y)$, with $x\in M$ and $y\in G$.
We call ``internal'' transformations those that leave the $x$ coordinates fixed, 
and ``spacetime'' transformations those that leave the $y$ coordinates fixed.
Generic transformations can be represented by pairs $(f,g)$ consisting
of a spacetime transformation $f$ and an internal transformation $g$.
As already noted, spacetime and internal transformations
do not commute in general, so we choose conventionally that the internal
transformation acts first.
The transformation $(f,g)$ maps the point $(x,y)\in P$
to the point $(f^{-1}(x),g^{-1}(x)y)$ and
the composition of two transformations is
$$
(f_2,g_2)\circ(f_1,g_1)=
(f_1\circ f_2,g_1\cdot( g_2\circ f_1^{-1}))\ ,\eqno(4)
$$
where $g_1\cdot g_2$ denotes multiplication in $G$.
One can now easily verify that the internal transformations are a normal
subgroup $\inv_*$ of $\inv$ but the spacetime transformations are not normal.
The quotient $\inv/\inv_*$ is isomorphic to $\diff$, the group of diffeomorphism of $M$, 
and $\inv$ is the semidirect product of $\inv_*$ and $\diff$.

In high energy experiments only the momenta and the charges of the incoming
and outgoing particles can be measured;
therefore the transformations that tend to the identity
at infinity are gauge invariances.
Thus $\gauge$ is the normal subgroup of $\inv$ consisting of
transformations that tend to the identity at infinity.
Note that the transformations of $\gauge$ have the property that
they leave the asymptotic behavior of the fields at infinity unchanged:
$\delta_\epsilon\phi\to 0$ and $\delta_\xi\phi\to 0$ for any field $\phi$.
In the group $\gauge$ there is a normal subgroup $\gauge_*=\gauge\cap\inv_*$
of internal gauge transformations, namely the internal transformations
that tend to the identity at infinity.
The quotient $\gauge/\gauge_*$ is the (normal) subgroup 
$\bar\diff\subset \diff$ of diffeomorphisms of $M$
that tend to the identity at infinity.
From (4) one sees that in the groups $\inv$ and $\gauge$, 
spacetime and internal transformations are always mixed in a nontrivial way.

Not so in the symmetry group. 
The transformations that leave the VEVs (3)
invariant are the ``rigid'' (constant) internal transformations 
and global Poincar\'e transformations.
These two groups commute and therefore the symmetry group is
$\sym^{(0)}=G\times$Poincar\'e
\footnote{($^3$)}{It is assumed that all the matter fields $\psi$ 
transforming linearly under $G$ have vanishing VEV.
If among them there were some Higgs field with a nontrivial VEV,
this would reduce the group $G$ to some subgroup $H$
and the discussion would go through with obvious modifications.}
.


As mentioned before, the choice of the VEV in (3) breaks the gauge invariance
of the theory. Does the result depend on the choice of classical fields
within the gauge equivalence class?
Suppose that we choose the VEVs $(A^{(1)}_\mu,g^{(1)}_{\mu\nu})$
to be gauge transforms of (3) with some gauge transformation $(f(x),g(x))$.
The symmetry group $\sym^{(1)}$ defined by $(A^{(1)}_\mu,g^{(1)}_{\mu\nu})$
is conjugate to $\sym^{(0)}$:
if $h$ is an isometry of $g^{(0)}$ (a Poincar\'e tranformation)
$f^{-1}\circ h\circ f$ is an isometry of $g^{(1)}=f^*g^{(0)}$,
and likewise for the internal trasformations.
Thus $\sym^{(1)}$ is isomorphic to $\sym^{(0)}$;
the symmetry group does not depend on the choice
of fields within a gauge equivalence class.

These arguments were based on consideration of the classical
configuration space of the theory, but
if we postulate that in addition to leaving the action invariant,
the transformations of $\inv$ also leave the functional integration
measure invariant, then at least formally they will also be invariances 
of the quantum theory. 
In particular the transformations of $\sym^{(0)}$
will give rise to symmetries of the $S$ matrix.
So, while our starting point and our line of reasoning were quite different, 
we have reached the same conclusion as the classic CM theorem.
In fact one could say that we have merely produced an example of application
of that theorem, were it not for the fact that these arguments can
now be generalized to other situations.

We could now consider non flat but asymptotically flat metrics and gauge fields;
to be even more general let us consider a different setup in which
the world is turned inside out.
Instead of thinking of the laboratory as a fixed reference frame at infinity,
as is natural in describing scattering experiments,
we think of it as a coordinate system in an infinitesimally small spacetime 
neighborhood of a point $\bar x$ in a gravitational field.
(Besides being of small extent in space, the laboratory is also supposed 
to operate for a very short time.)
We can give an idealized description of this observational setup as 
an orthonormal frame $\bar e$ in the tangent space at $\bar x$.
With this frame, one can directly measure the components of spacetime tensors
at $\bar x$.
We assume that the observer also has some apparatus
that can be used to measure the internal degrees of freedom 
of the matter fields $\psi$ at $\bar x$;
this apparatus can be described geometrically as a ``$G$--frame'' in internal space,
or equivalently a point $p_0\in P$ over $\bar x$.
Finally, we assume that the observer can measure the infinitesimal 
parallel transport at $\bar x$;
this is described geometrically by the horizontal lift $\tilde e$ of $\bar e$ 
at the point $p_0$
\footnote{($^4$)}{It is possible to drop this last condition without affecting
the final conclusions. Here we keep it so as to maintain a greater similarity
between the behavior of $g_{\mu\nu}$ and $A_\mu$.}.
Since fields with different values at $\bar x$ are physically distinguishable,
the variations given in (1) and (2) must vanish at $\bar x$ for gauge transformations.
This implies that the infinitesimal generators of gauge transformations
are parametrized by Lie algebra-valued functions $\epsilon$ and vector fields $\xi$ such that 
$$
\epsilon(\bar x)=0\ ; \ \  \partial_\mu\epsilon(\bar x)=0\ ;\ \ 
\xi^\nu(\bar x)=0\ ;\ \ \partial_\mu\xi^\nu(\bar x)=0\ .\eqno(5)
$$

This setup could be used to describe our position in a cosmological context.
The previously discussed case of scattering experiments in flat space 
could also be regarded as a special case of this construction
by thinking of $M$ as the conformal compactification of Minkowski
space and $\bar x$ as the point at infinity. 

Let us now assume that the VEV of the metric is not flat but the
VEV of the gauge field is still flat,
so that $P$ can again be described as the product $M\times G$.
There is no Poincar\'e group anymore, so we are definitely
outside the domain of applicability of the classic CM theorem.
Does this mean that we can have symmetry groups mixing internal
and spacetime transformations?

We observe that (5) implies {\it at the point} $\bar x$,
$$
\delta_\epsilon A^{(0)}_\mu=D^{(0)}_\mu\epsilon=0\ ;\ \ 
\delta_\xi A^{(0)}_\mu=\xi^\nu\partial_\nu A^{(0)}_\mu+A^{(0)}_\nu\partial_\mu\xi^\nu=0\ ;\ \
\delta_\xi g^{(0)}_{\mu\nu}=\nabla^{(0)}_\mu\xi_\nu+\nabla^{(0)}_\nu\xi_\mu=0\ ,\eqno(6)
$$
where a superscript $(0)$ over the covariant derivatives indicates that they are computed
with respect to the classical VEVs.
Thus, gauge transformations do not change the values of
$A^{(0)}_\mu(\bar x)$ and $g^{(0)}_{\mu\nu}(\bar x)$.
By contrast, the symmetry group consists of transformations that obey equations (6)
{\it everywhere on} $M$.
It will be the same as in the Minkowskian case, except for the replacement
of the Poincar\'e group by the isometry group of the metric $g^{(0)}$:
$\sym^{(0)}=G\times I(g^{(0)})$.
We conclude that when the VEV of the metric is not flat but the
VEV of the gauge field is flat, it is still true that
internal and spacetime symmetries (if any) do not mix.

Let us now consider more general situations when also the VEV of the
gauge field is nontrivial (we do not need to make any assumption
about the global topology).
In such cases there is no natural choice of trivialization for $P$
and as a result it is in general impossible to meaningfully split the
invariance group $\inv$ into spacetime and internal transformations.
The group $\inv$ is the group of automorphisms of $P$,
\ie\ diffeomorphisms of $P$ that map fibers into fibers
and commute with the right action of $G$ on $P$.
If an automorphism $u$ maps the fiber over $x$ to the fiber over $y$,
to $u$ there corresponds naturally a diffeomorphism $f$
which maps $x$ to $y$. The automorphisms $u$ for which $f$ is the
identity of $M$ are called the vertical automorphisms and form a normal subgroup $\inv_*$.
These transformations can meaningfully be called ``internal'',
and it is again true that $\inv/\inv_*=\diff$.
However now, unlike in previous examples, there is no natural way of realizing
$\diff$ as a subgroup of $\inv$.
In a given (local) trivialization one may call ``spacetime transformations''
those of the form $(x,y)\mapsto (f(x),y)$,
but in another trivialization the same transformation will no longer be of the same form,
and since there are no preferred trivializations there is no
preferred subgroup that one may invariantly call ``spacetime transformations''.
So, $\inv$ is not a semidirect product anymore.
Neither is the normal subgroup of gauge transformations $\gauge$ whose 
infinitesimal parameters satisfy equation (5).

The symmetry group is now more complicated to describe.
Since $\inv$ has a subgroup of internal transformations,
but no subgroup of spacetime transformations,
for a given field $A^{(0)}_\mu$ it is still meaningful to look for an algebra of
functions $\epsilon$ such that $\delta_\epsilon A^{(0)}_\mu=0$,
but it makes no longer sense to look for transformations such that $\delta_\xi A^{(0)}_\mu=0$. 
Instead one must in general look for pairs $(\epsilon,\xi)$ such that
$\delta_\epsilon A^{(0)}_\mu+\delta_\xi A^{(0)}_\mu=0$.
The problem of determining all gauge fields that are invariant in
this sense has been addressed in the literature [4,5].
We will not discuss this in detail but merely give a name $S(A^{(0)})$ 
to the group of transformations that leave $A_\mu^{(0)}$ invariant.

In general $S(A^{(0)})$ may have a subgroup  $S_*(A^{(0)})$ of internal symmetries
which generalizes the group of rigid internal transformations.
If we try to define a ``rigid'' action of a subgroup $K\subset G$ on $P$ by saying that
in a given (local) trivialization $(x,y)\mapsto (x,ky)$,
\footnote{($^5$)}{One cannot use the right action $(x,y)\mapsto (x,yk)$
because such transformations leave $\psi$ and $A_\mu$ invariant.}
the result is not well-defined in general, because such action
does not commute with the action of gauge transformations,
which is given by $(x,y)\mapsto (x,g(x)^{-1}y)$.
However, if the transition functions of the bundle $P$ have values in
$Z_G(K)$, the centralizer of $K$ in $G$, 
then we can identify a $Z_G(K)$-subbundle $Q\subset P$ and
we have a well-defined action of $K$ on $Q$.
Then, a gauge field has (internal) symmetry group $S_*(A^{(0)})=K$
if and only if it can be regarded as a connection in $Q$,
or equivalently its holonomy has values in $Z_G(K)$ [6].
This gives indeed $K=G$ if $A^{(0)}$ is flat and
if $G$ is abelian, $G$ is always a symmetry of any gauge field.

Returning to the general discussion, 
the symmetries of the theory must leave $A^{(0)}_\mu$ and $g^{(0)}_{\mu\nu}$ invariant,
so the condition (s3) reads
$$
D^{(0)}_\mu\epsilon+\xi^\nu\partial_\nu A^{(0)}_\mu+A^{(0)}_\nu\partial_\mu\xi^\nu=0\ ;\ \
\nabla^{(0)}_\mu\xi_\nu+\nabla^{(0)}_\nu\xi_\mu=0\ .\eqno(7)
$$
Condition (s2) says simply that the parameters $\epsilon$ and $\xi$ and their first 
derivatives must not all simultaneously vanish at $\bar x$.
The symmetry group is $\sym^{(0)}=I(g^{(0)})\cap S(A^{(0)})$, and
in general it consists of mixed internal and spacetime transformations.

Before discussing examples, let us return to the issue of gauge dependence:
the definition of the symmetry group that we have used so far
depends on the unphysical choice of representatives $(A^{(0)}_\mu,g^{(0)}_{\mu\nu})$.
The fact that the symmetry groups of different representatives are conjugate
subgroups in $\inv$, and hence isomorphic, should reassure us that we are not
being deceived by gauge illusions.
Still, to be completely sure, we now discuss an alternative definition of symmetry group 
that does not require the choice of a representative. 
We work in the general case when neither $A^{(0)}_\mu$ nor $g^{(0)}_{\mu\nu}$ are flat.
The groups $\inv$ and $\gauge$ are as before.
Since gauge related fields are physically indistinguishable,
a physical (though in practice not a very useful) description of the degrees of freedom of the
theory is by gauge equivalence classes of pairs $(A_\mu,g_{\mu\nu})$,
denoted $[A_\mu,g_{\mu\nu}]$.
If we denote $\conn$ the space of gauge fields and $\met$ the space of metrics,
the group $\inv$ acts on $\conn\times\met$ as in (1,2) and its normal subgroup
$\gauge$ acts on $\conn\times\met$ without fixed points,
so that $(\conn\times\met)/\gauge$ is a smooth infinite dimensional manifold [7,8].
Since $\gauge$ is normal in $\inv$, 
the action of $\inv$ on $\conn\times\met$ defines an action 
of $\inv/\gauge$ on $(\conn\times\met)/\gauge$.
The vacuum of the theory can be gauge invariantly described as a point 
$[A^{(0)}_\mu,g^{(0)}_{\mu\nu}]$ in $(\conn\times\met)/\gauge$
and according to condition (s3) the symmetry group of the theory is
the subgroup $\sym\subset\inv/\gauge$ that leaves the vacuum invariant.

This group can be described as follows.
Let $\inv^{(0)}\subset\inv$ be the subgroup of transformations $u$ that 
map a given representative pair of the vacuum $(A^{(0)}_\mu,g^{(0)}_{\mu\nu})$
to another pair in the same equivalence class.
This means that 
$$
(A^{(0)}_\mu,g^{(0)}_{\mu\nu})u=(A^{(0)}_\mu,g^{(0)}_{\mu\nu})u'\ ,\eqno(8)
$$
where $u'\in\gauge$. If $(A^{(1)}_\mu,g^{(1)}_{\mu\nu})$
is another representative in the same equivalence class, (8) implies that
$(A^{(1)}_\mu,g^{(1)}_{\mu\nu})u=(A^{(1)}_\mu,g^{(1)}_{\mu\nu})u''$
for some other gauge transformation $u''$ (this follows from the normality of $\gauge$).
Therefore the group $\inv^{(1)}$ coincides with $\inv^{(0)}$.
Thus, we can gauge invariantly characterize $\inv^{(0)}$ as the subgroup of 
$\inv$ that maps {\it any} representative of the vacuum into another
representative of the vacuum.
Now, obviously $\gauge\subset\inv^{(0)}$ is a normal subgroup 
but $\inv^{(0)}$ may contain some elements that are not in $\gauge$.
In fact equation (8) implies that $u=\bar u u'$ where $\bar u\in \sym^{(0)}$.
Then, the symmetry group of the theory is $\sym=\inv^{(0)}/\gauge$
and it is isomorphic to $\sym^{(0)}$, for any choice of representatives.

This can be seen a little more explicitly as follows.
All representatives of the vacuum have the same values at $\bar x$:
$(A^{(0)}_\mu(\bar x),g^{(0)}_{\mu\nu}(\bar x))$
and all infinitesimal gauge transformations satisfy (7) at the point $\bar x$.
The Lie algebra of $\sym$ consists of infinitesimal transformations
that satisfy (7) at $\bar x$ without satisfying (5).
Such transformations are parametrized by the values of $\epsilon$, $\xi^\mu$ and 
their first derivatives at $\bar x$.
Given any representative of the vacuum $(A^{(0)}_\mu,g^{(0)}_{\mu\nu})$
if the equations (7) admit solutions, such solutions are
uniquely characterized by the initial data provided by the 
values of $\epsilon$, $\xi^\mu$ and their first derivatives at $\bar x$.
\footnote{($^6$)}{Let us stress that equations (7) cannot in general
be solved for arbitrary such initial data. In particular
the derivatives $\partial_\mu\epsilon$ are not free parameters.}
.
Thus, the Lie algebra of the group $\sym$ is isomorphic to the Lie algebra
of the group $\sym^{(0)}$ discussed earlier,
for any fixed pair $(A^{(0)}_\mu,g^{(0)}_{\mu\nu})$
chosen in the gauge equivalence class of the vacuum.
We see that the alternative, gauge invariant definition of the vacuum
gives the same result as the one based on the background field method.

Let us summarize the main conclusions of this discussion.
We have carefully distinguished between generic invariances of the action,
gauge invariances and symmetries in a gauge theory coupled to gravity.
In the gauge group, spacetime and internal transformations are always mixed
in the sense that they do not form commuting subgroups.
If the vacuum can be described by (the gauge equivalence class of)
classical fields $A^{(0)}_\mu$ and $g^{(0)}_{\mu\nu}$, the symmetry group
is $\sym=I(g^{(0)})\cap S(A^{(0)})$, where 
$I(g^{(0)})$ is the isometry group of $g^{(0)}_{\mu\nu}$ and
$S(A^{(0)})$ is the symmetry group of $A^{(0)}_\mu$.
This result is gauge invariant.
It holds also for asymptotically flat vacuum states,
if we interpret $\bar x$ as the point at infinity, 
as in the first example.

The symmetry group may contain a subgroup $S_*(A^{(0)})\subset G$
of internal symmetry transformations.
Such internal rigid transformations can only be defined in special cases.
For example, if $A^{(0)}$ is flat $S_*(A^{(0)})=G$ is the group
of ``rigid'' internal transformations.
In general if $A^{(0)}$ is not flat one cannot define these transformations.
A physical consequence of this, which has been known since long,
is the impossibility of defining the nonabelian
charges in the background of a monopole [9,10].
The preceding discussion shows that this kind of phenomenon is more general
and is not limited to gauge fields with nontrivial topology.

If one considers transformations which are not the identity of $M$,
and if the VEV of the gauge field is not flat,
in general $\inv$ has no subgroup that can be identified
as spacetime transformations and the transformations in
$\sym$ are actually mixtures of internal and spacetime transformations.
Examples of this phenomenon have also been known since long in soliton physics.
For example nonabelian monopoles do not have separate symmetries
for spatial rotations and internal transformations;
the only symmetries are combinations of these transformations
and the conserved Noether charges are the sum of
angular momentum, spin and isospin generators [11,12].
Note that when the metric is flat, as is the case in these examples, any nonflat gauge field will
break Poincar\'e invariance, so this mixing of internal and spacetime symmetries
does not conflict with the original CM theorem.

These investigations have been stimulated in part by recent discussions of unified theories.
One way of achieving a unification of gravity and gauge interactions 
is to treat the Lorentz (gravitational) connection and the Yang--Mills gauge field 
(for some group $G$) as components of a connection of a larger unifying group [13-17].
There has been some debate about the way in which
such theories avoid conflict with the CM theorem.
For instance, in some models, due to the presence of a cosmological constant, 
the ground state of the theory would be de Sitter space.
The CM theorem does not hold in de Sitter space, so, it has been argued, 
such theories would allow mixing between internal and spacetime symmetries.
To clarify this point, recall that the order parameter for such unification
is some generalized version of the vierbein.
The theory can be in various phases, depending on the VEV of this order parameter.
In the ``broken'' phase in which gravity is separated from other gauge interactions,
the VEV of the metric is nonzero; let us assume for the sake of argument
that it is de Sitter space, and that the VEV of the $G$ gauge field is flat.
The classic CM theorem cannot be applied, but the preceding discussion shows that
there would still be no mixing between the internal and spacetime (de Sitter) symmetries.
So, the cosmological constant is irrelevant for this issue.
On the other hand in the ``fully symmetric'' phase the VEV of the metric would be zero.
This is a different, ``topological'' state of the theory that we have not even considered here.
As discussed in [14], the CM theorem does not forbid the mixing of internal and spacetime symmetries 
in a topological phase, nor does any other argument of the type given here.
\bigskip

\goodbreak
{\bf Acknowledgements.} This work is partially supported by the
INFN-MIT Bruno Rossi exchange program.
I wish to thank A. Ashtekar, J. Goldstone, R. Jackiw and A. Manohar
for discussions.

\bigskip 
\centerline{\bf References}
\item{[1]} S. Coleman and J. Mandula, Phys. Rev. {\bf 159} 1251 (1967).
\item{[2]} S. Elitzur, Phys. Rev. D 12, 3978 - 3982 (1975).
\item{[3]} L. 'O Raifeartaigh, in Lecture Notes in Physics 379 (1991).
\item{[4]} J. Harnad, S. Shnider and L. Vinet, J. Math. Phys. {\bf 21} 2719 (1980).
\item{[5]} P. Forgacs and N. Manton, Comm. Math. Phys. {\bf 72} 15 (1980).
\item{[6]} P.A. Horvathy and J.W. Rawnsley, J. Math. Phys. {\bf 27} 982 (1986).
\item{[7]} A.E. Fischer, in ``Relativity'', ed. M. Carmeli, S.Fickler and L.Witten, Plenum,
New York (1967); D.G. Ebin, A.M.S. Bull {\bf 74} 1001 (1968).
\item{[8]} P.K. Mitter and C.M. Viallet, Comm. Math. Phys. {\bf 79} 457 (1981).
\item{[9]} P. Nelson and A. Manohar, Phys. Rev. Lett. {\bf 50} 943 (1983). 
\item{[10]} A.P. Balachandran, G. Marmo, N. Mukunda, J.S. Nilsson, E.C.G. Sudarshan 
and F. Zaccaria, Phys. Rev. Lett. {\bf 50} 1553 (1983).
\item{[11]} R. Jackiw and C. Rebbi, Phys. Rev. Lett. {\bf 36} 1116 (1976).
\item{[12]} P. Hasenfratz and G. 't Hooft, Phys. Rev. Lett. {\bf 36} 1119 (1976).
\item{[13]} R. Percacci, Phys. Lett.  {\bf B 144} 37 (1984); Nucl. Phys.  {\bf B 353} 271 (1991), e-Print: arXiv:0712.3545 [hep-th].
\item{[14]} F. Nesti and R. Percacci, J. Phys. A: Math. Theor. {\bf 41} 075405 (2008), arXiv:0706.3307 [hep-th].
\item{[15]} Stephon H.S. Alexander, e-Print: arXiv:0706.4481 [hep-th].
\item{[16]} A.Garrett Lisi, e-Print: arXiv:0711.0770 [hep-th].
\item{[17]} L. Smolin, e-Print: arXiv:0712.0977 [hep-th].

\vfil
\eject
\bye